\pdfoutput=1

\documentclass[11pt,twoside,a4paper,cmspaper,final,collab]{cms-tdr}
\def\svnVersion{493495P}\def\cmsCernNoTag{CERN-EP-2018-292}\def\cmsCernDate{\today}\def\cmsMessage{Published in Physical Review Letters as \href{http://dx.doi.org/10.1103/PhysRevLett.122.121803}{\doi{10.1103/PhysRevLett.122.121803.}}}

\begin{document}\cmsNoteHeader{HIG-17-030}

\hyphenation{had-ron-i-za-tion}
\hyphenation{cal-or-i-me-ter}
\hyphenation{de-vices}
\RCS$HeadURL: svn+ssh://svn.cern.ch/reps/tdr2/papers/HIG-17-030/trunk/HIG-17-030.tex $
\RCS$Id: HIG-17-030.tex 493482 2019-04-05 12:47:01Z gortona $
\newlength\cmsFigWidth
\ifthenelse{\boolean{cms@external}}{\setlength\cmsFigWidth{0.98\columnwidth}}{\setlength\cmsFigWidth{.9\linewidth}}
\ifthenelse{\boolean{cms@external}}{\providecommand{\cmsLeft}{upper\xspace}}{\providecommand{\cmsLeft}{left\xspace}}
\ifthenelse{\boolean{cms@external}}{\providecommand{\cmsRight}{lower\xspace}}{\providecommand{\cmsRight}{right\xspace}}
\ifthenelse{\boolean{cms@external}}{\providecommand{\NA}{\ensuremath{\cdots}\xspace}}{\providecommand{\NA}{\ensuremath{\text{---}}\xspace}}
\ifthenelse{\boolean{cms@external}}{\providecommand{\CL}{C.L.\xspace}}{\providecommand{\CL}{CL\xspace}}
\ifthenelse{\boolean{cms@external}}{\providecommand{\CLnp}{C.L\xspace}}{\providecommand{\CLnp}{CL\xspace}}
\ifthenelse{\boolean{cms@external}}
{\providecommand{\suppMaterial}{the supplemental material~\cite{supplemental}}}
{\providecommand{\suppMaterial}{Appendix~\ref{app:suppMat}}}
\newcommand{\bbtt}{\ensuremath{\PQb\PQb\PGt\PGt}\xspace} 
\newcommand{\bbbb}{\ensuremath{\PQb\PQb\PQb\PQb}\xspace} 
\newcommand{\bbgg}{\ensuremath{\PQb\PQb\PGg\PGg}\xspace} 
\newcommand{\bbvv}{\ensuremath{\PQb\PQb\text{VV}}\xspace} 
\newcommand{\HH}{\ensuremath{{\PH\PH}}\xspace}
\newcommand{\lambdahhh}{\ensuremath{\lambda_{\PH\hspace{-.08em}\PH\hspace{-.08em}\PH}}\xspace} 

\cmsNoteHeader{HIG-17-030}
\title{Combination of searches for Higgs boson pair production in proton-proton collisions at \texorpdfstring{$\sqrt{s} = 13\TeV$}{sqrt(s) = 13 TeV}}

\date{\today}

\abstract{
This Letter describes a search for Higgs boson pair production using the combined results from four final states: \bbgg, \bbtt, \bbbb, and \bbvv, where V represents a $\PW$ or \PZ boson. The search is performed using data collected in 2016 by the CMS experiment from LHC proton-proton collisions at $\sqrt{s} = 13\TeV$, corresponding to an integrated luminosity of 35.9\fbinv.
Limits are set on the Higgs boson pair production cross section. A 95\% confidence level observed (expected) upper limit on the nonresonant production cross section is set at 22.2\,(12.8) times the standard model value.
A search for narrow resonances decaying to Higgs boson pairs is also performed in the mass range 250--3000\GeV. No evidence for a signal is observed, and upper limits are set on the resonance production cross section.
}

\hypersetup{
pdfauthor={CMS Collaboration},
pdftitle={Combination of searches for Higgs boson pair production in proton-proton collisions at sqrt(s) = 13 TeV},
pdfsubject={CMS},
pdfkeywords={CMS, physics, Higgs}}

\maketitle
The discovery of the Higgs boson ($\PH$) by the ATLAS and CMS Collaborations~\cite{ATLASdiscovery,CMSdiscovery,Chatrchyan:2013lba} was a major step in the understanding of the mechanism of electroweak symmetry breaking.
The Higgs boson mass was jointly determined by the two experiments using the CERN LHC Run 1 data to be $m_\PH=125.09\pm0.24\GeV$~\cite{Aad:2015zhl}
and recently by CMS using partial Run 2 data with even better accuracy, $m_\PH=125.26\pm0.21\GeV$~\cite{Sirunyan:2017exp}. With the Higgs boson mass known with a precision better than 0.2\%, the structure of the Higgs scalar field potential and the Higgs boson self-couplings are precisely predicted in the standard model (SM).
The current measurements of the properties of the Higgs boson are compatible with the SM predictions~\cite{Khachatryan:2016vau,Sirunyan:2018koj}. The measurement of the Higgs boson self-coupling provides an independent test of the SM and verification that the Higgs mechanism is truly responsible for the electroweak symmetry breaking by giving access to the shape of the Higgs scalar field potential~\cite{Baglio:2012np}. Access to the Higgs boson trilinear coupling can be obtained by measuring the production of pairs of Higgs bosons (\HH) at the LHC. At the same time, several theories predict heavy resonances that decay into \HH~\cite{Binoth:1996au,Schabinger:2005ei,Patt:2006fw,Branco20121,Fayet:1974pd,Fayet:1977yc,PhysRevD.76.036006, Fitzpatrick:2007qr}.
Studies of pair production of Higgs bosons, each of which can decay in different channels, allow one to probe different regions of the anomalous couplings space and of the resonant invariant mass spectrum. A combination of different channels is therefore needed to obtain the best possible sensitivity for the \HH production.

The \HH SM production cross section, which partly depends on the Higgs trilinear coupling, was computed at next-to-next-to-leading order in quantum chromodynamics (QCD), including next-to-next-to-leading-logarithmic corrections and finite top quark mass effects at next-to-leading order. Its value is
$\sigma_{\HH} = 33.53^{+4.3\%}_{-6.0\%} (\text{QCD scale}) \pm 5.9\% (\text{other})\unit{fb}$ in proton-proton ($\Pp\Pp$) collisions at 13\TeV for a Higgs boson mass of 125\GeV~\cite{deFlorian:2016spz,Borowka:2016ehy,Borowka:2016ypz,deFlorian:2015moa,Degrassi:2016vss}, where ``other'' includes contributions from parton distribution function (PDF) uncertainties evaluated using the PDF4LHC recommendations~\cite{Botje:2011sn,Alekhin:2011sk,Ball:2014uwa}, strong coupling constant \alpS dependence, and top quark mass effects.

Physics beyond the SM (BSM) can significantly modify the cross section and the kinematic properties of nonresonant Higgs boson pair production. In order to provide model independent constraints on these effects, we introduce an effective field theory (EFT) Lagrangian that extends the SM with dimension-6 operators~\cite{Goertz:2014qta}. This approach results in five anomalous Higgs boson couplings relevant for \HH production:
the $\PH$ coupling to the top quark, $y_\PQt$; the trilinear coupling, \lambdahhh; and three additional couplings, denoted by $\text{c}_2$, $\text{c}_{2\Pg}$, and $\text{c}_\Pg$, that represent, respectively, the interactions of a top quark pair with a Higgs boson pair, of a gluon pair with a Higgs boson pair, and of a gluon pair with a single \PH~\cite{deFlorian:2016spz}. We define $k_\lambda=\lambdahhh/\lambda_\text{SM}$ and $k_\PQt=y_\PQt/y^\text{SM}_\PQt$.
Since a full five-dimensional scan of all possible coupling combinations would be computationally excessive,
a clustering strategy~\cite{Carvalho:2016rys} has been developed to group together possible combinations of coupling values that present similar kinematic properties. Twelve clusters have been identified, in addition to the SM and the $\lambdahhh=0$ scenarios. Within each cluster, a representative point in the EFT space that we refer to as a benchmark is selected. Each benchmark thus represents a possible modification of the \HH signal yield and kinematic distributions due to BSM effects.

In the searches discussed in the Letter, the resonant signal is represented by either a CP-even particle of spin-0 (radion) or spin-2 (graviton) with a width that is much smaller than the detector resolution for the whole range under study.

The ATLAS and CMS Collaborations performed studies of Higgs boson pair production at $\sqrt{s}=8$\TeV~\cite{PhysRevD.92.092004,Khachatryan:2016sey,Sirunyan:2017tqo}. Limits on nonresonant \HH production were set by the CMS Collaboration in the \bbgg and \bbtt final states. Here and in the rest of the text the indication of the charge of the decay products is omitted for simplicity of notation. The combination of those searches allowed the CMS Collaboration to set an upper limit on \HH production at 43 times the SM expectation~\cite{Sirunyan:2017tqo}.
Using the data collected in 2016 at $\sqrt{s}=13$\TeV, the CMS and ATLAS Collaborations performed searches in the \bbgg~\cite{CMS:2017ihs,Aaboud:2018ftw}, \bbtt~\cite{Sirunyan:2017djm,Aaboud:2018sfw}, and \bbbb~\cite{CMS:2017xxp,Sirunyan:2017isc,bbbbnonresB2G,bbbbnonres,Aaboud:2018knk} final states, with the CMS Collaboration having results in the \bbvv~\cite{Sirunyan:2017guj} channel as well, where $\text{V}$ denotes either a $\PW$ or a $\PZ$ boson that decays leptonically. All four channels studied at CMS are included in this combination.

The CMS apparatus features a superconducting solenoid of 6\unit{m} internal diameter, providing a magnetic field of 3.8\unit{T}. Within the solenoid volume are a silicon pixel and strip tracker, a lead tungstate crystal electromagnetic calorimeter, and a brass and scintillator hadron calorimeter, each composed of a barrel and two end cap sections. Forward calorimeters extend the pseudorapidity coverage provided by the barrel and end cap detectors. Muons are detected in gas-ionization chambers embedded in the steel flux-return yoke outside the solenoid. Events of interest are selected using a two-tiered trigger system~\cite{Khachatryan:2016bia}.
The particle-flow algorithm~\cite{CMS-PRF-14-001} aims to reconstruct and identify each individual particle in an event with an optimized combination of information from the various elements of the CMS detector.
Dedicated \PQb tagging algorithms~\cite{btagging} are used to identify jets originating from \PQb quarks (\PQb jets).
A more detailed description of the CMS detector, together with a definition of the coordinate system used and the relevant kinematic variables, can be found in Ref.~\cite{Chatrchyan:2008zzk}.

With the current data, the \bbgg~\cite{CMS:2017ihs} analysis is the most sensitive to SM Higgs boson pair production in CMS.
Despite a low branching fraction  ($\mathcal{B}(\HH\to \PQb\PQb\PGg\PGg)=0.26\%$), the analysis profits from a very small background contribution, thanks to the excellent diphoton mass resolution of the CMS experiment for this channel (${\approx}1.6\GeV$~\cite{CMS:2017ihs}). To exploit this feature, the analysis relies on a 2D fit of the $\PH\to\PGg\PGg$ and $\PH\to \PQb\PQb$ invariant mass distributions, where the background coming from the $\text{N}\PGg+\text{jets}$ $(N=\left\{0,1,2,..\right\})$ continuum is estimated from the mass sidebands. Other background contributions can arise from single $\PH$ production and constitute up to a third of the total background in the nonresonant searches. They are modeled from Monte Carlo (MC) simulations, including gluon fusion, vector boson fusion, V\PH, $\ttbar\PH$, and $\PQb\bar{\PQb}\PH$ production processes.
In the nonresonant searches events are classified into low- and high-mass categories according to the \HH pair reduced mass $\tilde{M}_\text{X} = M_{\text{jj}\PGg\PGg} - M_\text{jj}- M_{\PGg\PGg} + 250$\GeV, which increases the sensitivity to \HH production. A further categorization based on the purity of the events is applied to both resonant and nonresonant searches. The signal purity is estimated by means of a boosted decision tree (BDT) built from the \PQb tagging probability of each jet, the angles between the products of the \HH system decay, and the transverse momentum of each \PH candidate.
The category most sensitive to nonresonant production is the one with the highest signal purity and reduced mass, with a SM signal over background ratio (S/B) of ${\approx}5$\%. The low reduced mass categories enhance sensitivity to large $k_\lambda$ values.
In the resonant search, two different BDTs are trained for resonance masses $m_\text{X}$ higher or lower than 600\GeV.

The \bbtt analysis~\cite{Sirunyan:2017djm} combines a relatively low background contamination with a relatively large branching fraction (7.3\%), for a final S/B of ${\approx}0.4$\% in the most sensitive category.
At least one isolated hadronically decaying $\PGt$ lepton must be present in the event, together with a second isolated lepton that is oppositely charged.
The second lepton can be either an electron or a muon (semileptonic final state) or another hadronically decaying $\Pgt$ lepton (fully hadronic final state).
Events are categorized according to the number of \PQb-tagged jets (one or two) in the event. Events with a boosted \PH jet candidate are assigned to a dedicated boosted category.
A BDT discriminant based on the kinematic differences between the \HH and $\ttbar$ processes is used in the semileptonic channels in order to reduce the large amount of $\ttbar$ background.
A kinematic fit~\cite{svfit} is performed to reconstruct the Higgs boson mass.
For the nonresonant searches, the so-called `stransverse mass' \mTii~\cite{Barr2014308,Sirunyan:2017djm,LESTER199999} is found to provide the best separation between the background and the signal~\cite{Barr2014308}.
For the resonant searches, a further kinematic fit, detailed in Ref.~\cite{Khachatryan:2015tha}, is performed to reconstruct the most probable mass of the resonance.
The largest background contributions come from Drell--Yan, $\ttbar$, processes and events comprised uniquely of jets produced via the strong interaction (QCD multijet). Additional background sources considered are V\PH, VV, and \PW+jets.
The Drell--Yan background is estimated from a leading order simulation with a scale factor obtained from data in a \PZ boson enriched control region to account for higher-order corrections. The QCD multijet background is estimated using three control regions with different lepton charge (same-sign region) and isolation requirements. All other background contributions are estimated from MC simulations.

The \bbbb final state has the highest \HH decay branching fraction (33.6\%). Three optimized resonance searches are performed targeting  resonances of different masses. For a resonance mass below 0.7\TeV, four $\PQb$-tagged jets~\cite{CMS:2017xxp} are required in the final state.
For high-mass searches ($m_\text{X}>1.2\TeV$), selected events must contain two ``\PH jet candidates," where each \PH jet candidate is a jet associated with a boosted $\PH$ decaying into a $\PQb\PAQb$ pair merged into a single jet~\cite{Sirunyan:2017isc}. This candidate reconstruction uses jet substructure and jet flavor tagging techniques~\cite{PhysRevLett.100.242001}.
The search sensitivity in the intermediate mass range, 0.7--1.2\TeV, is improved by considering both the final state with four \PQb jets and the case with one Higgs jet candidate and two \PQb jets~\cite{bbbbnonresB2G}.
A dedicated nonresonant \HH search is performed in the four \PQb jet final state~\cite{bbbbnonres}. Triggers using \PQb tagging and jet substructure techniques are used to collect these events. The sensitivity of the analysis is improved using a BDT that employs jet-related and \HH decay kinematic variables and global event properties. Under SM assumptions, the S/B can be as high as 1\% in the most sensitive BDT bins. Further sensitivity to BSM nonresonant \HH production, which often results in boosted topologies, is obtained adding the final states with one or two \PH jet candidates~\cite{bbbbnonresB2G}.
The dominant background comes from QCD multijet events and it is estimated using data sideband regions~\cite{CMS:2017xxp,Sirunyan:2017isc,bbbbnonresB2G} and a hemisphere mixing technique~\cite{bbbbnonres}. The residual background is dominated by $\ttbar$ events and is estimated using simulation. Such events correspond to 10, 15, and 1\% of the total background in the low-, intermediate-, and high-mass ranges, respectively, for resonant searches and  5 (15)\% for nonresonant searches in the final states with four \PQb-tagged jets (with \PH jet candidates).

The \bbvv analysis~\cite{Sirunyan:2017guj} includes the $\HH\to \PQb\PQb \PW\PW\to \PQb\PQb \ell\nu \ell\nu$ and the $\HH\to \PQb\PQb \PZ\PZ\to \PQb\PQb\ell\ell\nu\nu$ processes, for a total branching fraction $\mathcal{B}(\HH\to \PQb\PQb\ell\nu\ell\nu)=2.72\%$ when all possible lepton flavor decays are considered (including the \PGt leptonic decays).
This channel has large backgrounds, which are predominantly from $\ttbar$ and Drell--Yan processes. All background sources are estimated from MC simulations with the exception of Drell--Yan, which is estimated from data. In order to improve the rejection of these large backgrounds, a deep neural network (DNN) discriminant technique has been implemented. Two separate parametrized DNNs have been trained, one for the resonant and one for the nonresonant search. The first is parametrized according to the mass of the resonance, while the second depends on the parameters $k_\lambda$ and $k_\PQt$. For each nonresonant benchmark, the closest point in the $(k_\lambda, k_\PQt)$ plane, according to the statistical measure defined in Ref.~\cite{Carvalho:2016rys}, has been used to define the DNN parameters. An S/B ratio of ${\approx}0.18\%$ is obtained in the most sensitive DNN region for SM nonresonant production.

For both the resonant and nonresonant searches, likelihood fits are performed using the statistical toolkits, \textsc{RooFit}~\cite{Verkerke:2003ir} and \textsc{RooStats}~\cite{Moneta:2010pm}. The signal strength ($\mu$), defined as the ratio between the observed and expected signal rates, is estimated with its corresponding confidence interval via the profile likelihood ratio test statistic~\cite{CMS-NOTE-2011-005}. The latter depends on the signal strength as well as on the nuisance parameters, which account for various experimental and theoretical uncertainties. The reference value for the expected signal strength is chosen as the SM gluon fusion \HH production cross section (33.53\unit{fb}) in nonresonant searches.
The likelihood fits are performed with respect to the observed data or a data set constructed using $\mu=1$ for assessing expected results using the asymptotic approximation~\cite{Cowan:2010st,Junk:2016jag}. Limits are set at 95\% confidence level (\CL) using the \CLs criterion~\cite{CLS1,CLS2}.
For all measurements, the $\PH$ mass is fixed at $m_\PH=125\GeV$, and branching fractions are assumed to be equal to the SM predictions.
When investigating signal models corresponding to the shape benchmarks, the single \PH production cross sections are all assumed to have their SM values.

When combining results, the systematic uncertainties from various sources are accounted for as follows.
A polynomial interpolation between alternative shape variations is used to model systematic effects on the shape of the discriminant variables.
Lepton and photon reconstruction and identification efficiencies and energy scale corrections are assumed to be fully correlated across the channels and analysis categories that use the same objects.
The uncertainties in the jet energy scale corrections are divided into multiple sources. The effect of each source on the rate and shape of the final observable for different processes is considered in the \bbvv analysis and is treated as fully correlated with the normalization effects in the \bbtt and \bbgg channels. In these latter channels, the shape effects from the individual sources have been verified to be negligible, and only the cumulative effects on the shape are considered. They are assumed to be correlated with the normalization and shape variation from the \bbbb analysis. Jet energy resolution effects are negligible in the \bbtt channel and assumed to be fully correlated between the \bbvv and the \bbbb channels. The effects of jet energy scale and resolution are included in the functional form for the shape used in the \bbgg channel and are assumed to be uncorrelated with the other channels.
Uncertainties related to the \PQb tagging are dominated by the modeling of heavy flavor production when measuring data/MC scale factors, and are considered correlated
across the \bbtt, \bbgg, resonant \bbbb, and \bbvv channels. The nonresonant \bbbb analysis uses a different \PQb tagging method and is considered uncorrelated.

An integrated luminosity uncertainty of 2.5\%~\cite{lumipaper} is applied in a fully correlated way to all channels and all processes estimated from simulation.
Different uncertainties are applied to background processes that are estimated from data.
The uncertainties in the total cross sections of the common background processes are assumed to be correlated across all channels, and are of the order of 5\% for the most relevant ones (single top, \ttbar, V\PH).
The uncertainty in the same-sign to opposite-sign candidate ratio in the \bbtt channel is propagated to the estimation of the multijet background, and the uncertainties in the scale factors applied to the $\PZ/\PGg^*\to \ell\ell$ background estimation to correct for higher-order effects are also taken into account. A normalization uncertainty
is considered in the \bbbb nonresonant search to account for residual biases in the hemisphere mixing technique.
Some background estimates in the \bbtt, \bbvv, and nonresonant \bbbb analyses have non-negligible statistical uncertainties due to the limited number of events passing the selection in data control regions or simulated data samples.
These are taken into account by allowing each bin in each template shape to fluctuate independently according to a Poisson distribution. These uncertainties are assumed to be uncorrelated across bins in the individual template shapes.
The nonresonant signal uncertainties include contributions coming from variations in the renormalization and factorization scales, these amount to $^{+4.3\%}_{-6.0\%}$~\cite{deFlorian:2016spz,deFlorian:2015moa} of the nonresonant signal cross section. Other theoretical uncertainties such as those in \alpS, PDFs, and finite top quark mass effects at next-to-next-to-leading order result in a further 5.9\% uncertainty~\cite{Degrassi:2016vss,Borowka:2016ehy,Borowka:2016ypz}. These effects are assumed to be fully correlated across the different channels.
The \alpS, PDFs, and scale variation effects are also included for single $\PH$ background contributions. These uncertainties are considered fully correlated for the V\PH production in the \bbtt, \bbvv, and \bbgg channels.
The uncertainty in the branching fraction of the Higgs boson to $\PQb\PQb$~\cite{deFlorian:2016spz} is also assumed to be fully correlated across all channels.

The event yield in data is small, so the statistical uncertainties are much larger than the systematic ones. Those with the largest effect on the final result are the statistical fluctuations in the yield in the most sensitive bins of the BDT and the overall background normalization in the \bbbb channel, the  hadronically decaying $\tau$ lepton energy scale effects in the \bbtt analysis, and the uncertainty in the functional form used to model the signal shape in the \bbgg channel. These effects are as large as 10 (5)\% for the \bbbb and \bbtt (\bbgg) uncertainties. Due to its lower overall sensitivity, the systematic uncertainties affecting the \bbvv analysis have little effect on the combined result. The largest sources of systematic uncertainty for this channel arise from the uncertainties in the $\ttbar$ cross section, electron identification efficiency, and \PQb tagging efficiency.

\begin{figure}
	\centering
	\includegraphics[width=\cmsFigWidth]{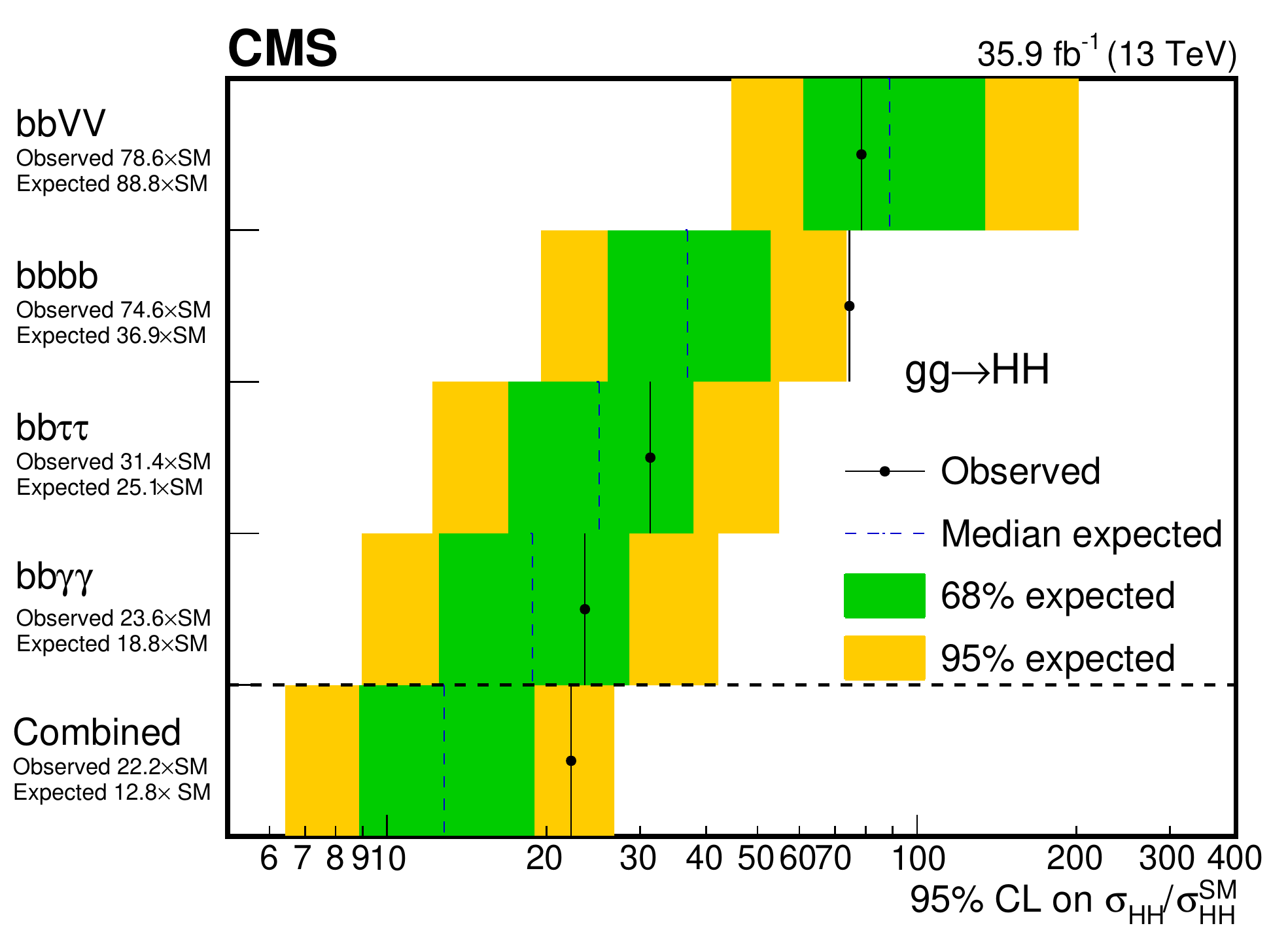}
	\caption[]{The 95\% \CL upper limits on the signal strength $\mu=\sigma_{\HH}/\sigma_{\HH}^\text{SM}$.
	The inner (green) band and the outer (yellow) band indicate the regions containing 68 and 95\%, respectively, of the limits on $\mu$ expected under the background-only hypothesis.}
	\label{fig:smlim}
\end{figure}

\begin{figure}
	\centering
		\includegraphics[width=\cmsFigWidth]{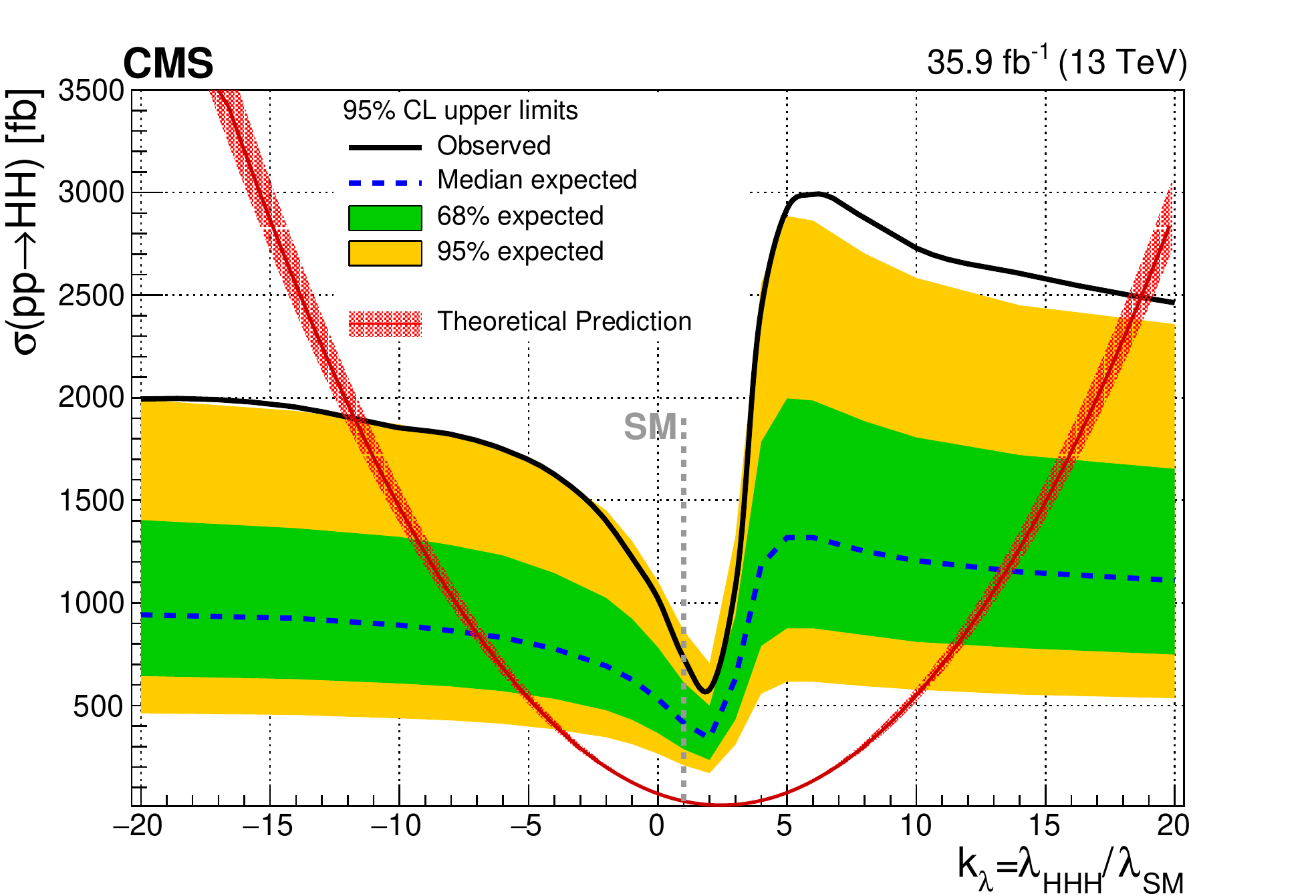}
		\caption{Expected and observed 95\% \CL upper limits on the \HH production cross section as a function of the $k_\lambda$ parameter. The inner (green) band and the outer (yellow) band indicate the regions containing 68 and 95\%, respectively, of the limits on the \HH cross section expected under the background-only hypothesis.
			The red band shows the expected theoretical~\cite{Carvalho:2016rys} cross section expectations and its uncertainty. All other couplings and EFT parameters are set to their SM values.}
		\label{fig:klscan}
	\end{figure}

With all the correlations across the channels included, the observed and expected limits at 95\% confidence level on the nonresonant \HH production signal strength are measured to be 22.2 and 12.8 times the SM expectations, respectively. They are shown in Fig.~\ref{fig:smlim} for the individual channels and their combination. Small excesses, compatible with statistical fluctuations, are observed in the \bbbb, \bbtt, and \bbgg final states and result in a small excess in the combined result.
A scan is performed for different values of the $k_\lambda$ parameter, while keeping all other EFT parameters fixed at their SM values. The value of $k_\lambda$ affects both the expected cross section and the \HH decay kinematics. For each value, these differences are fully simulated and considered in the scan. Resulting limits are reported in Fig. \ref{fig:klscan}.
The exclusion limit as a function of $k_\lambda$ closely follows the features of the \HH production cross section and \HH invariant mass distribution $M_{\HH}$~\cite{Frederix:2014hta}, which are sculpted by the interference between the \HH production via the trilinear Higgs coupling and the emission of an \HH pair from a top quark loop. The minimum at $k_\lambda=2.46$ corresponds to the maximum negative interference between the two diagrams, which results in a minimum of the cross section but at the same time enhances the relative importance of tails in the $M_{\HH}$ distribution.
The maximum at $k_\lambda{\approx}5$ is due to the softness of the $M_{\HH}$ spectrum for such values of the trilinear coupling, causing a larger fraction of events to fall outside experimental acceptance. As $\abs{k_\lambda}$ increases, the production via the trilinear Higgs coupling becomes dominant and the limit asymptotically approaches the same value for both $k_\lambda\ll-10$ and $k_\lambda\gg10$.
This is reflected in the observed exclusion limit as well, where the significance of the small observed excess is relatively less important in the more sensitive small $k_\lambda$ region than at large values of $k_\lambda$. When fixing all the other EFT parameters to their SM values, the $k_\lambda$ parameter is observed (expected) to be constrained to the range $-11.8<k_\lambda<18.8$ ($-7.1<k_\lambda<13.6$) at 95\% \CLnp.
The observed exclusions for the different EFT benchmarks~\cite{Carvalho:2016rys} are in the range of 100--3000\unit{fb}, and can be seen in \suppMaterial. A small excess, similar to that observed at the SM value, is present across most of the phase space with the exception of the more boosted topologies.
	\begin{figure}
		\centering
		\includegraphics[width=\cmsFigWidth]{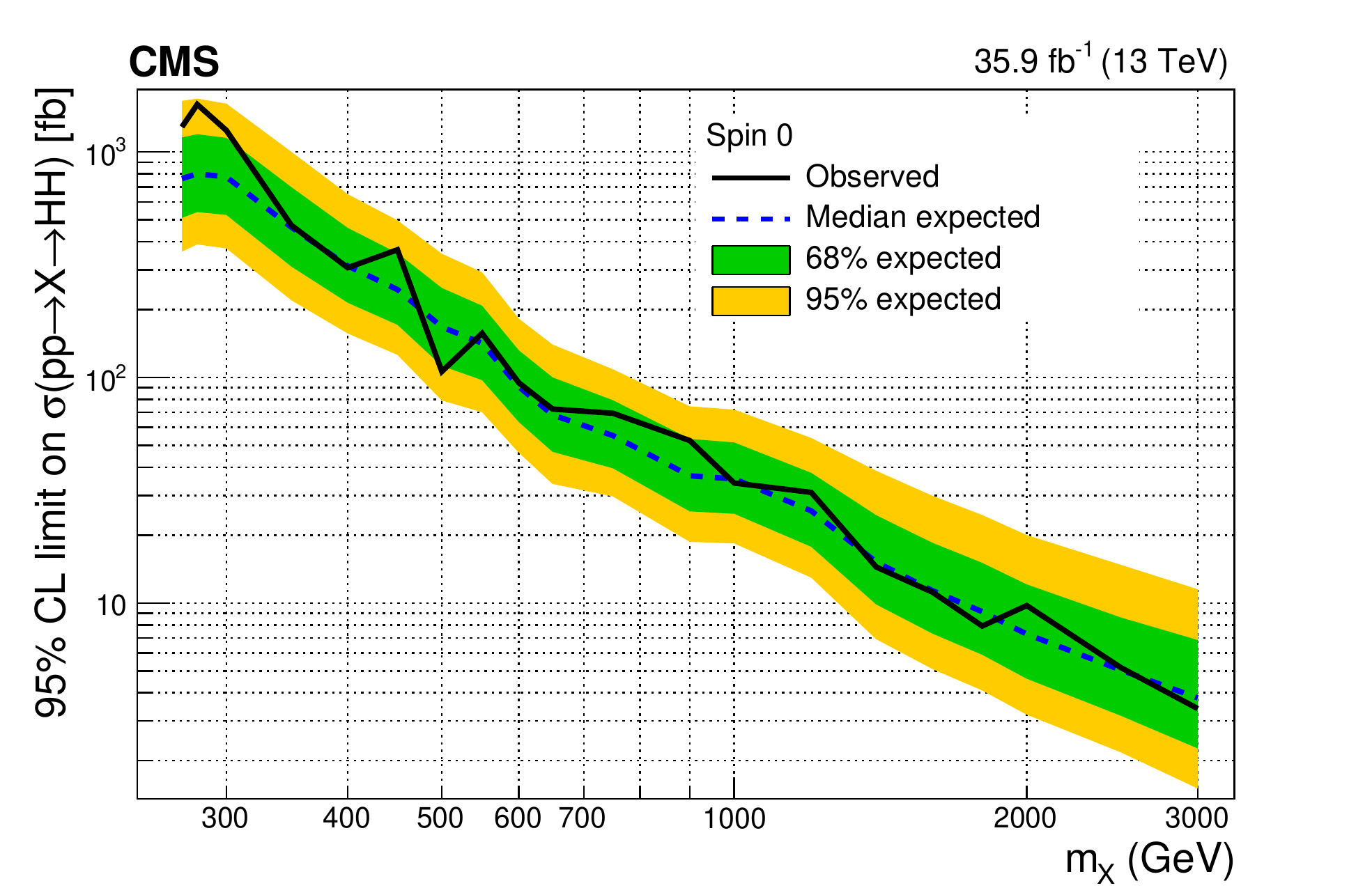}
		\caption{Expected (dashed) and observed (solid line) 95\% \CL exclusion limits on the production of a narrow, spin zero resonance (X) decaying into a pair of Higgs bosons. The inner (green) band and the outer (yellow) band indicate the regions containing 68 and 95\%, respectively, of the limits on the \HH cross section expected under the background-only hypothesis.}
		\label{fig:resonant}
\end{figure}

The resonant search is performed in the range of masses from 250 to 3000\GeV. Under the hypothesis of a narrow-width resonance, no significant excess is found across the whole range for either a spin-0 or a spin-2 resonance. The results of the combined resonant search are shown in Fig. \ref{fig:resonant} for the spin-0 model, and in \suppMaterial{} for the spin-2 case.

In summary, a combination of searches for nonresonant and resonant Higgs boson pair production has been presented. The combination includes the \bbgg, \bbtt, \bbbb, and \bbvv channels, where V represents a $\PW$ or \PZ boson, using a data sample collected in proton-proton collisions at $\sqrt{s} = 13\TeV$, which corresponds to an integrated luminosity of 35.9\fbinv.
Upper limits at 95\% confidence level (\CL) on the Higgs boson pair production cross section are obtained. For the nonresonant production mechanism, the observed (expected) 95\% \CL corresponds to 22.2\,(12.8) times the theoretical prediction for the standard model cross section.
 An effective field theory framework is used to parametrize the cross section as a function of anomalous couplings of the Higgs boson. When varying only the ratio $k_\lambda$ between the Higgs trilinear coupling \lambdahhh and its standard model expectation, values of $k_\lambda$ in the region $-11.8<k_\lambda<18.8$ ($-7.1<k_\lambda<13.6$) are still allowed by the observed (expected) data.
 For the resonant production mechanism, upper exclusion limits at 95\% \CL are obtained for the production of a narrow resonance with mass ranging from 250 to 3000\GeV.
These results represent both the most sensitive and most comprehensive study of double Higgs production at the LHC to date.

\begin{acknowledgments}
We congratulate our colleagues in the CERN accelerator departments for the excellent performance of the LHC and thank the technical and administrative staffs at CERN and at other CMS institutes for their contributions to the success of the CMS effort. In addition, we gratefully acknowledge the computing centers and personnel of the Worldwide LHC Computing Grid for delivering so effectively the computing infrastructure essential to our analyses. Finally, we acknowledge the enduring support for the construction and operation of the LHC and the CMS detector provided by the following funding agencies: BMBWF and FWF (Austria); FNRS and FWO (Belgium); CNPq, CAPES, FAPERJ, FAPERGS, and FAPESP (Brazil); MES (Bulgaria); CERN; CAS, MoST, and NSFC (China); COLCIENCIAS (Colombia); MSES and CSF (Croatia); RPF (Cyprus); SENESCYT (Ecuador); MoER, ERC IUT, and ERDF (Estonia); Academy of Finland, MEC, and HIP (Finland); CEA and CNRS/IN2P3 (France); BMBF, DFG, and HGF (Germany); GSRT (Greece); NKFIA (Hungary); DAE and DST (India); IPM (Iran); SFI (Ireland); INFN (Italy); MSIP and NRF (Republic of Korea); MES (Latvia); LAS (Lithuania); MOE and UM (Malaysia); BUAP, CINVESTAV, CONACYT, LNS, SEP, and UASLP-FAI (Mexico); MOS (Montenegro); MBIE (New Zealand); PAEC (Pakistan); MSHE and NSC (Poland); FCT (Portugal); JINR (Dubna); MON, RosAtom, RAS, RFBR, and NRC KI (Russia); MESTD (Serbia); SEIDI, CPAN, PCTI, and FEDER (Spain); MOSTR (Sri Lanka); Swiss Funding Agencies (Switzerland); MST (Taipei); ThEPCenter, IPST, STAR, and NSTDA (Thailand); TUBITAK and TAEK (Turkey); NASU and SFFR (Ukraine); STFC (United Kingdom); DOE and NSF (USA).
\end{acknowledgments}

\bibliography{auto_generated}
\ifthenelse{\boolean{cms@external}}{}{
	\clearpage
	\appendix
\section{Supplemental material \label{app:suppMat}}

Figure \ref{fig:benchallchan} shows the 95\% \CL upper limit on the \HH production cross section for each benchmark topology, the standard model and the $k_\lambda=1$ case and for each final state. As can be seen in the plot, different benchmark shapes result in very different sensitivities for a given analysis. The combined limits alone are shown in Fig. \ref{fig:benchmarks}.

The sensitivity of each analysis to the SM \HH production is shown in Fig. \ref{fig:smlim_wrun1}, together with the 8\TeV CMS result.

The exclusion limits for a spin-2 narrow width resonance are shown in Fig. \ref{fig:resonant2}.

\begin{sidewaysfigure}
		\includegraphics[width=\linewidth]{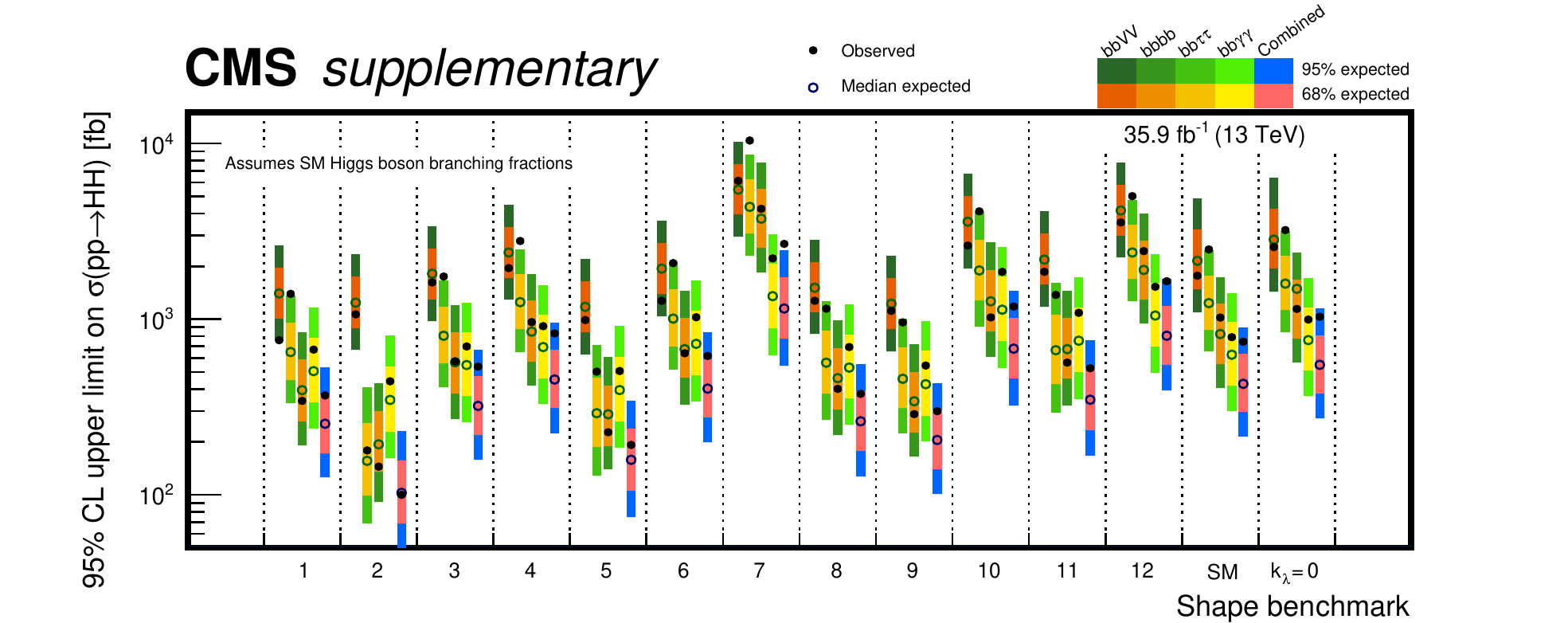}
		\caption[]{The 95\% \CL exclusion limits on the nonresonant Higgs boson pair production cross sections for different EFT benchmark topologies (bins 1 to 12) as defined in Ref.~\cite{Carvalho:2016rys}. Each benchmark represents a possible modification in the \HH signal yield and kinematic distributions due to BSM effects. The last two bins show the 95\% \CL exclusion limits for the SM and for the $k_\lambda$= 0 scenarios. Limits are shown for each of the four final states separately and for the combination.}
		\label{fig:benchallchan}
\end{sidewaysfigure}
\begin{figure}
	\centering
	\includegraphics[width=\linewidth]{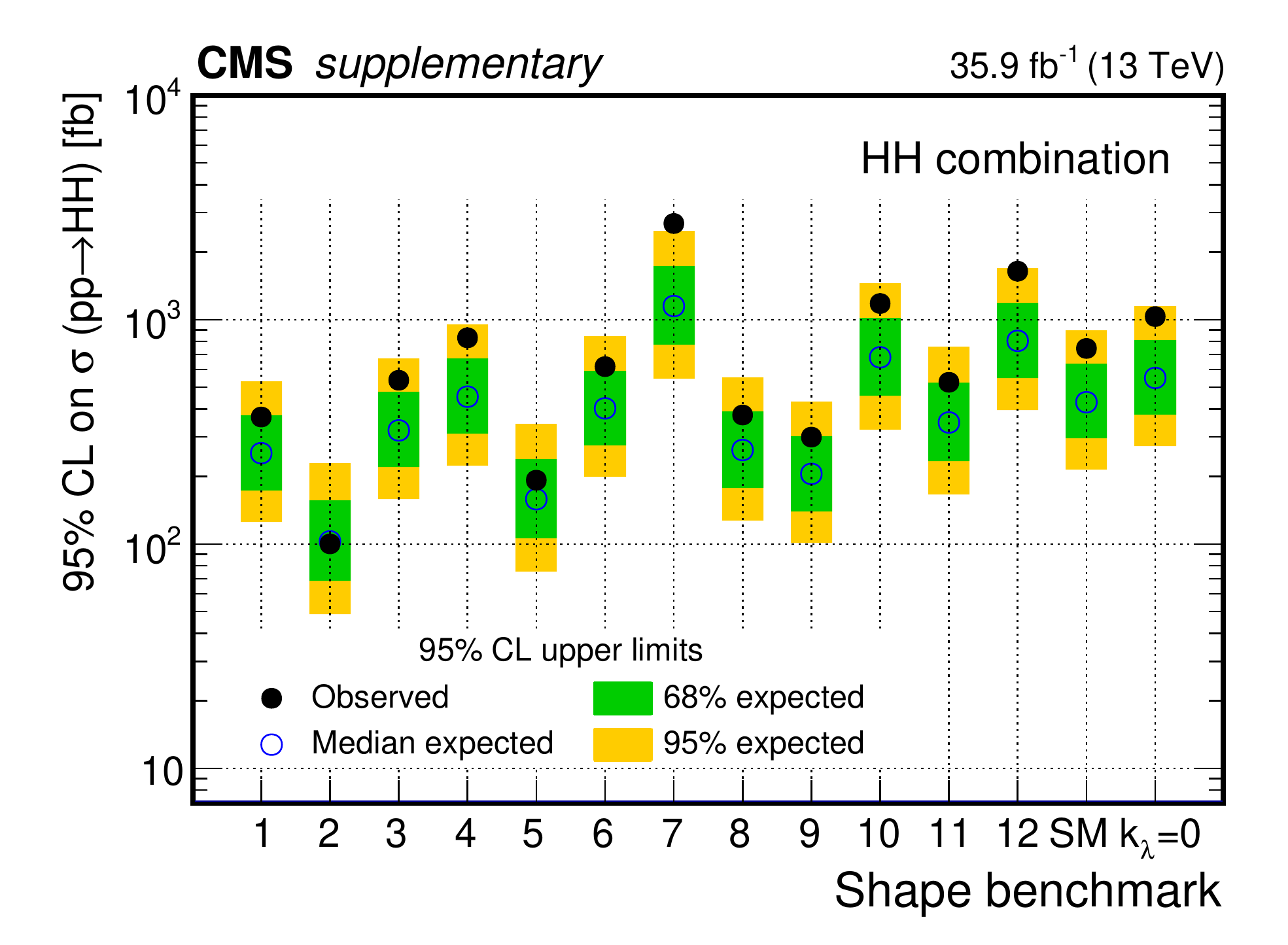}
	\caption[]{The 95\% \CL exclusion limits on the nonresonant Higgs boson pair production cross sections for different EFT benchmark topologies. Each bin of the histogram shows the limit for a different benchmark (bins 1 to 12) as defined in Ref.~\cite{Carvalho:2016rys},  for SM production and for $k_\lambda=0$ production (last 2 bins). Each benchmark represents a possible modification in the \HH signal yield and kinematic distributions due to BSM effects.}
	\label{fig:benchmarks}
\end{figure}
\begin{figure}
	\centering
	\includegraphics[width=\linewidth]{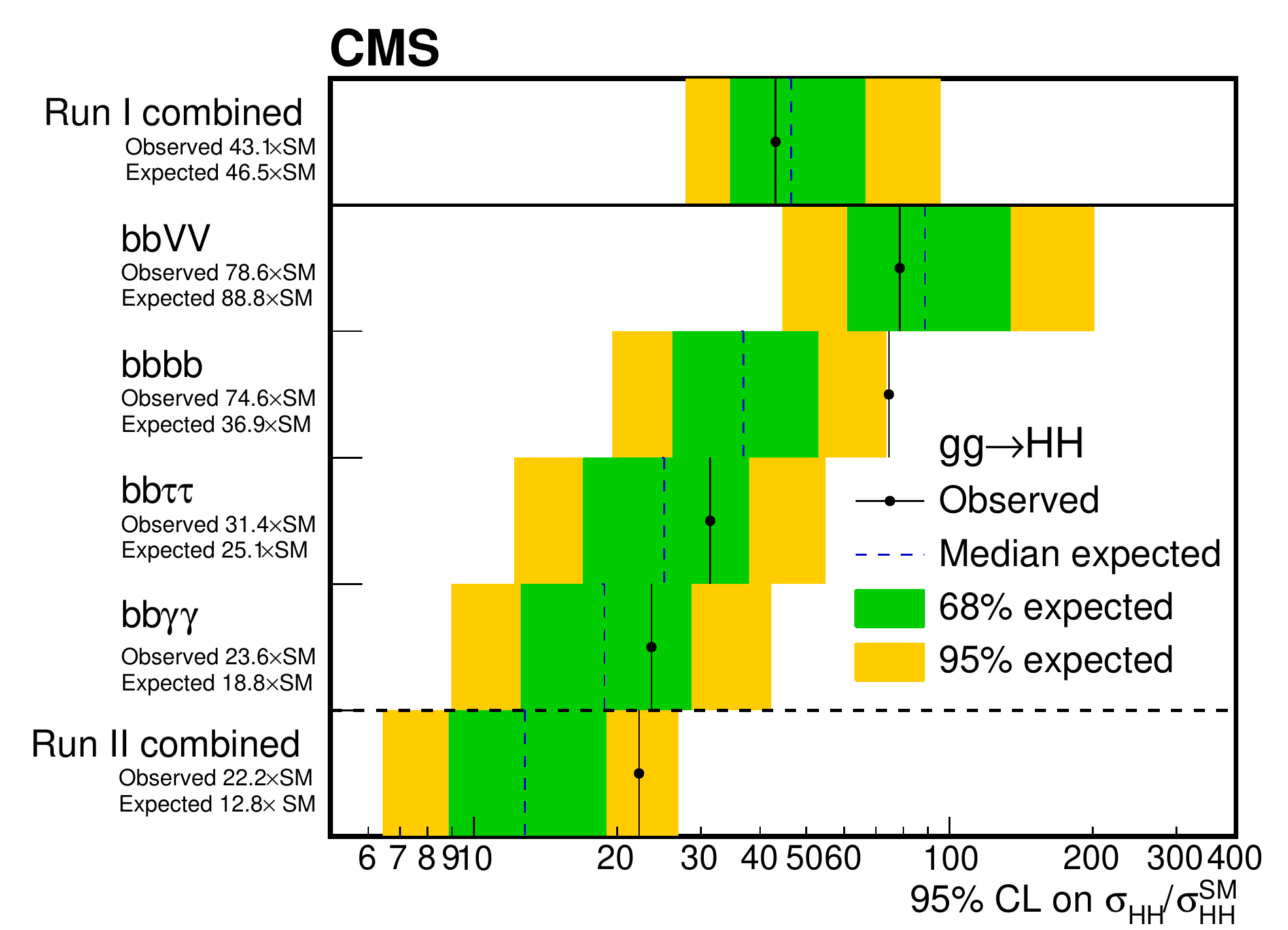}
	\caption[]{The 95\% \CL exclusion limits on the SM nonresonant Higgs boson pair production cross section for different channels. The inner (green) band and the outer (yellow) band indicate the regions containing 68 and 95\%, respectively, of the distribution of limits expected under the background-only hypothesis. The uppermost bin reports the combined limit obtained at $\sqrt{s}=8$\TeV.}
	
	\label{fig:smlim_wrun1}
\end{figure}

\begin{figure}
	\centering
	\includegraphics[width=\linewidth]{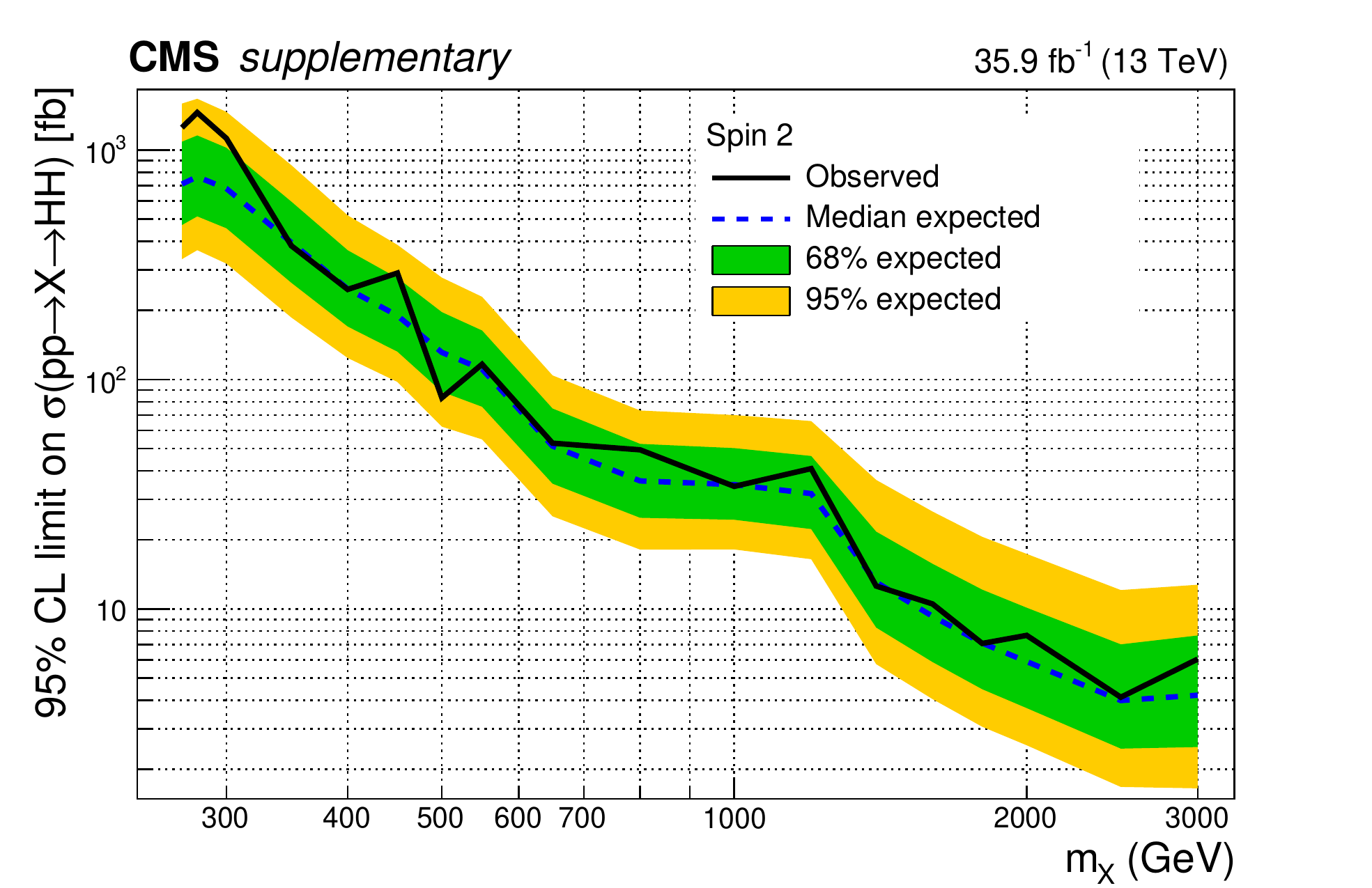}
	\caption{Expected (dashed) and observed (solid line) 95\% \CL exclusion limits on the production of a narrow, spin-2 resonance decaying into a pair of Higgs bosons. The inner (green) band and the outer (yellow) band indicate the regions containing 68 and 95\%, respectively, of the distribution of limits expected under the background-only hypothesis.}
	\label{fig:resonant2}
\end{figure}

}

\cleardoublepage \section{The CMS Collaboration \label{app:collab}}\begin{sloppypar}\hyphenpenalty=5000\widowpenalty=500\clubpenalty=5000\vskip\cmsinstskip
\textbf{Yerevan Physics Institute, Yerevan, Armenia}\\*[0pt]
A.M.~Sirunyan, A.~Tumasyan
\vskip\cmsinstskip
\textbf{Institut f\"{u}r Hochenergiephysik, Wien, Austria}\\*[0pt]
W.~Adam, F.~Ambrogi, E.~Asilar, T.~Bergauer, J.~Brandstetter, M.~Dragicevic, J.~Er\"{o}, A.~Escalante~Del~Valle, M.~Flechl, R.~Fr\"{u}hwirth\cmsAuthorMark{1}, V.M.~Ghete, J.~Hrubec, M.~Jeitler\cmsAuthorMark{1}, N.~Krammer, I.~Kr\"{a}tschmer, D.~Liko, T.~Madlener, I.~Mikulec, N.~Rad, H.~Rohringer, J.~Schieck\cmsAuthorMark{1}, R.~Sch\"{o}fbeck, M.~Spanring, D.~Spitzbart, A.~Taurok, W.~Waltenberger, J.~Wittmann, C.-E.~Wulz\cmsAuthorMark{1}, M.~Zarucki
% ...
\vskip\cmsinstskip
% ...
1:  Also at Vienna University of Technology, Vienna, Austria\\
% ...
\end{sloppypar}
\end{document}